\documentclass{article}

\usepackage{amsfonts}
\usepackage{float}
\usepackage{amsmath}
\usepackage{amssymb}
\usepackage{amsthm}
\usepackage{graphicx}
\usepackage{multirow}
\usepackage{color}
\usepackage{ulem}
\usepackage{setspace}
\usepackage{caption}
\usepackage{subcaption}
\usepackage{rotating}
\usepackage{threeparttable}
\usepackage{url}

{\catcode `\@=11 \global\let\AddToReset=\@addtoreset}
\AddToReset{equation}{section}

\AddToReset{Theorem}{section}

\newtheorem{corollary}{Corollary}[section]
\newtheorem{lem}{Lemma}[section]
\newtheorem{theorem}{Theorem}[section]

\newtheorem{definition}{Definition}[section]

\theoremstyle{remark}
\newtheorem{rem}{Remark}[section]

\newcommand{\cT}{{\cal T}}
\newcommand{\cU}{{\cal U}}

\def\bc{\begin{center}}
\def\bd{\begin{description}}
\def\be{\begin{enumerate}}
\def\ec{\end{center}}
\def\ed{\end{description}}
\def\edt{\end{document}}
\def\ee{\end{enumerate}}
\def\ben{\begin{equation}}
\def\benn{\begin{equation*}}
\def\een{\end{equation}}
\def\eenn{\end{equation*}}
\def\benr{\begin{eqnarray}}
\def\eenr{\end{eqnarray}}
\def\benrr{\begin{eqnarray*}}
\def\eenrr{\end{eqnarray*}}

\def\edt{\end{document}}

\def\G{\Gamma}

\def\iny{\infty}

\def\lel{\label}

\def\noi{\noindent}

\def\si{\sigma}

\def\sti{\sum_{i=1}^n}

\def\th{\theta}

\def\wh{\widehat}

\def\R{{\mathbb R}}

\def\bc{\begin{center}}
\def\ec{\end{center}}

\title{A Goodness of Fit Test for Non-Gaussian Distributions with Unknown Location and Scale Parameters}
\author{Jiwoong Kim\\
University of South Florida}

\begin{document}
\maketitle
\noindent\\
\centerline{\Large{\textbf{Abstract}}}
\noindent\\
This paper studies computational aspects of an asymptotically distribution-free goodness-of-fit test for non-Gaussian distributions based on the Khmaladze martingale transformation when the location and scale parameters of the distribution are unknown. On top of that, we propose another goodness-of-fit test better than existing one in terms of a statistical power. Simulation studies demonstrate that the proposed test compares favorably with the existing test.\\
\noi

\section{Introduction}
Scientists have been trying to explain data from natural phenomena by employing statistical methodologies. Of those methodologies, a goodness-of-fit (GOF) test which specifies the underlying distribution of an observed random sample has always been of great interest to them. To illustrate the importance of the GOF test, consider a mechanical valve that is used in heart surgery. Proper functioning of a mechanical valve is critical in that a heart will, otherwise, fail to pump blood, thereby causing death to a patient. From this perspective, being fully informed about a lifetime of a mechanical valve is of paramount importance to both a patient and a physician since a mechanical valve should be replaced in a timely manner before it becomes worn-out and broken. One way to establish a standard for replacement is to construct a statistical confidence interval for the lifetime, which requires information on the probability distribution of the lifetime of the valve. At this crucial juncture, the GOF test can play a pivotal role in specifying the distribution of the lifetime of the mechanical valve by screening out false distributions. There are many other examples which serve to show the importance of the GOF test. Well reflecting its importance, there is now a colossal body of literature on the GOF test for a normal distribution. Crucially, however, research on the GOF test for non-Gaussian distributions has not been active as much as research on its real application, to say the least. It is against this backdrop of a rarity of a GOF test for other non-normal distributions that we started this study.

Khmaladze (1981) proposed the asymptotic-distribution-free (ADF) method through martingale transformation which is referred to as Khmaladze martingale transformation (KMT) test. Shedding light on solving an issue of the specification of unknown parameters which frequently arises in hypothesis test for distributions, Khmaladze (1981) demonstrated that the resulting test statistics is aymptotically distributed as the Brownian motion. Extending the application of the KMT test to the setup of regression models, Khmaladze and Koul (2004) demonstrated the same conclusion holds when the test statistic is based on the residuals. For other works on the KMT test, see Koul and Zhu (2015), Koul and Sakhanenko (2005), Tsigroshvili (1998), etc.

The KMT test is known to be versatile in that it can test for a wide range of distributions. Even though the versatility of the KMT test definitely merits further study, research on the KMT test is, however, still relatively undeveloped due to complexities which are immanent in the computation of its test statistic. In an effort to remove this computational issue which impedes the successful application of the KMT test, Kim (2020) proposed a fast and efficient algorithm for implementing the KMT test. The principles discussed in Kim (2020) cannot, however, be taken as absolute and general, and hence, have a certain drawback in that only a normal distribution was examined. Due to this reason, his approach hitherto has received but scant attention. While findings in Kim (2020) were confined to a normal distribution, the main idea of the proposed algorithm -- referred to as ``the strategy of integration-in-advance" -- can be commonly applied to other distributions which belong to the location-scale family. The present study is designed to examine whether computational aspects about the KMT test -- which were discussed in Kim (2020) -- are still valid for non-Gaussian distributions, and hence, elaborate on the idea that the KMT test can be applied to testing for other distributions belonging to the location-scale family. The main contribution of this paper is to propose the necessary and  sufficient condition for applicability of the KMT test for non-Gaussian distributions. On top of that, we propose a better-powered KMT test which is a slight modified from the original test.

The rest of this article is organized as follows. Section \ref{sec:discussion} provides a brief summary of the KMT test and states the main result of this article. Section \ref{sec:simulation} provides simulation studies which demonstrates that the new proposed KMT test empirically has a larger statistical power than the original KMT test. An \texttt{R} package \texttt{GofKmt} used for the simulation studies is available from Comprehensive \texttt{R} Archive Network (CRAN) at \url{https://cran.r-project.org/web/packages/GofKmt/index.html}. In the remainder of this article, we use the following notations; for any function $f:\R\rightarrow \R$ which is differentiable, $f'$ or $\dot{f}$ denotes the first derivative of $f$; for any function $g:\R\rightarrow \R$, $g^{2}(x)$ denotes a square of $g(x)$. For a differentiable function $h:\R\rightarrow \R$, $\dot{h}^{2}(x)$ will, therefore, denote $\{h'(x)\}^{2}$. For an $n\times n$ matrix $\mathbf{M}$, $\mathbf{M}^{-1}$ will denote its inverse matrix. For an $n\times 1$ vector $\mathbf{a}$, $\mathbf{a}'$ will denote its transpose vector.

\section{KMT test}\lel{sec:discussion}
In the past few decades, several statisticians conducted research on the KMT method and its application to various relevant problems: see, e.g., Koul and Zhu (2015), Koul and Sakhanenko (2005), and Tsigroshvili (1998). Among several references for the KMT test, Fan and Koul (2006, Chapter 9) is reckoned to be the best compendium for those who do not have some background knowledge about it: they explicated the theoretical perspective of the KMT test in detail. With some notations being borrowed from them, the summary of the KMT test in this section has a root in their work. Let $X_{1},X_{2},...,X_{n}$ be independent and identically distributed (i.i.d.) random variables where $F_{\theta}$ is a common distribution function (d.f.),\,\,and $f_{\theta}$ is an absolutely continuous density function with $\boldsymbol{\theta}:=(\mu,\sigma)'\in \R^{2}$ being a vector of location and scale parameters. For example, the density and distribution functions of the logistic random variable are written as
\benn
f_{\theta}(x) = \frac{e^{-(x-\mu)/\si}}{\si(1+e^{-(x-\mu)/\si})^{2}}\,\,\textrm{ and }\,\, F_{\th}(x) = \frac{1}{1+e^{-(x-\mu)\si}}
\eenn
while those of the Cauchy random variable are
\benn
f_{\theta}(x) = \frac{1}{\pi\sigma[1+((x-\mu)/\sigma)^2]}\,\,\textrm{ and }\,\, F_{\th}(x) = \frac{1}{2}+\frac{1}{\pi}\arctan((x-\mu)/\sigma).
\eenn
Let $\boldsymbol{\theta}_{0}:=(0,1)'\in \R^{2}$ and consider a GOF problem of testing for a distribution of a location-scale family
\ben\lel{eq:1}
H_{0}: F_{\theta}(x)=F_{\theta_{0}}\big((x-\mu)/\sigma\big)\textrm{ for all }x\in\R, \textrm{ vs. }H_{a}:\,H_{0}\textrm{ is not true}.
\een
To conserve the space, we drop $\theta_{0}$ in the subindex and let $F(x)$ and $f(x)$ denote $F_{\theta_{0}}$ and $f_{\theta_{0}}$, respectively. In this study, we need the following assumptions:
\be
\item[(\textbf{F.1})] $f$ is uniformly and absolutely continuous with $f>0$ almost everywhere (a.e.).
\item[(\textbf{F.2})]$\sup_{x\in\R}|xf(x)|<\iny$.
\item[(\textbf{F.3})] $f$ has a.e.\,\,derivative $\dot{f}$ satisfying
    \benn
    \int \left[ \left(\frac{\dot{f}(x)}{f(x)}\right)^{2} +\left( 1+ \frac{x\dot{f}(x)}{f(x)}\right)^{2}\right]\,dF(x)<\iny.
    \eenn
\ee
For more details of (\textbf{F.1})-(\textbf{F.3}), see Fan and Koul (2006, p.\,187). Furthermore, we assume that
\be
\item[(\textbf{E})] There exist consistent estimators of $\mu$ and $\sigma$, e.g., the maximum likelihood estimators.
\ee

For $x\in \R$ and $t=F(x)$, define
\benr\lel{eq:Gamma}
&&\phi(x):= -\dot{f}(x)/f(x),\quad \boldsymbol{\ell}(x):=(1, \phi(x), x\phi(x)-1 )',\\
&&p(t):= f(F^{-1}(t)),\quad q(t):= F^{-1}(t)f(F^{-1}(t)),\nonumber\\
&&\G_{t}=  \left(
                 \begin{array}{ccc}
                   1-t & p(t) & q(t) \\
                   p(t) & \int_{t}^{1}\dot{p}^{2}(u)du & \int_{t}^{1}\dot{p}(u)\dot{q}(u)du \\
                   q(t) & \int_{t}^{1}\dot{p}(u)\dot{q}(u)du & \int_{t}^{1}\dot{q}^{2}(u)du \\
                 \end{array}
               \right).\nonumber
\eenr
\begin{definition}
A given distribution $F$ is \textit{Khmaladze-transformable} if $\G_{t}$ is nonsingular for all $t\in (0,1)$. We call the set of all Khmaladze-transformable distributions \textit{the Khmaladze-transformable family}.
\end{definition}
\begin{rem}
As shown in Kim (2022), a normal distribution is, for example, Khmaladze-transformable. In this article, we shall show that the logistic and Cauchy distributions also belong to the Khmaladze-transformable family: see, e.g., Lemmas \ref{lem:c1}.
\end{rem}

Next, define $\wh{X}_i:= (X_{(i)}-\wh{\mu}_n)/\wh{\si}_n$
where $\wh{\mu}_n$ and $\wh{\si}_n$ are consistent estimators of $\mu$ and $\sigma$ -- e.g., the maximum likelihood (ML) estimators -- under the null hypothesis and $X_{(1)}\leq\cdots\leq X_{(n)}$ are order statistics of $X_{i}'s$. Observe that $\wh{X}_{1}\le \wh{X}_{2} \le \cdots \le \wh{X}_{n}$. The martingale transformed process based on $\{\wh{X}_i:\,\,i=1,2,...,n\}$ is defined as
\ben\lel{eq:MTP}
\small{\mathcal{U}_{n}(z)}:= \normalsize{n^{-1/2}\sti \left\{  I(\wh{X}_{i}\leq z) - \int_{-\iny}^{z\wedge \wh{X}_{i}} \boldsymbol{\ell}(\wh{X}_{i})'\G_{F(x)}^{-1}\boldsymbol{\ell}(x)f(x)\,\,dx   \right\} }
\een
for $z\in \R$ where $I(\cdot)$ is an indicator function. The original KMT test statistic is defined as
\ben\lel{eq:teststat}
\mathcal{T} := \sup_{z\in \R}|\mathcal{U}_{n}(z)|.
\een
As Khmaladze (1981) demonstrated,  $\mathcal{U}_{n}(z)$ converges to the Brownian motion in distribution, which implies the KMT test is indeed ADF: see, also, Khmaladze and Koul (2004). Motivated by the idea of the ADF property, we propose slightly modified KMT test which is defined as
\ben\lel{eq:TM}
\mathcal{T}^{M}= \sup_{z\in \R}\,\mathcal{U}_{n}(z)- \inf_{z\in \R}\,\mathcal{U}_{n}(z).
\een
As shown later, the same result (e.g., in Theorem \ref{thm:1}) holds for both $\mathcal{T}$ and $\cT^{M}$, and hence, we focus on the original KMT test and defer discussion about the modified test to the end of this section: it will be shown, therein, that the modified test yields a better power than the original test.

With $t=F(x)$, define $v_{0}$, $v_{1}$, and $v_{2}$ as
\benn
v_{0}(x):=\int_{t}^{1} \dot{p}(u)^{2}du,\,\,
v_{1}(x):=\int_{t}^{1} \dot{p}(u)\dot{q}(u)du,\,\,
v_{2}(x):=\int_{t}^{1} \dot{q}(u)^{2}du,
\eenn
where $v_{0}$, $v_{1}$, and $v_{2}$ are the (2,2)th, (2,3)th, and (3,3)th entries of $\Gamma_{t}$. Note that $v_{i}$'s can be rewritten as
\benrr
v_{0}(x)&=&\int_{x}^{\iny}\phi(y)^{2}f(y)dy,\\ v_{1}(x)&=&\int_{x}^{\iny}\phi(y)(y\phi(y)-1)f(y)dy,\\ v_{2}(x)&=&\int_{x}^{\iny}(y\phi(y)-1)^{2}f(y)dy,
\eenrr
and hence, the assumption (\textbf{F.3}) implies $v_{i}(x)<\iny,\,i=0,1,2$ for all $x\in\R$ while $v_{i}(x)>0,\,i=0,2$ follows from the assumption (\textbf{F.1}). On the other hand, $v_{1}$'s of the logistic and Cauchy distributions are still strictly positive, whereas that of Gumbel distribution is not: see, e.g, Lemma \ref{lem:vi}.

Recall that computation of $\cT$ necessitates the inverse of $\Gamma_{t}$. Being entangled in many subsequent equations which are integral to the computation of $\Gamma_{t}^{-1}$, e.g., appearing in denominators of the equations which arise in the middle of the computation, the characteristics of $v_{i}$'s (especially, $v_{0}$) are, hence, the most crucial for the proof of the nonsingularity of $\Gamma_{t}$. In case of the Cauchy distribution, $v_{i}$'s are rewritten as
\benrr
v_{0}(x)&=&\frac{1}{2}(1/2-\arctan(x)/\pi)+\frac{x(1-x^2)}{2\pi(1+x^2)^2},\,\,v_{1}(x)= \frac{x^2}{\pi(1+x^2)^2},\\
v_{2}(x)&=&\frac{1}{2}(1/2-\arctan(x)/\pi)-\frac{x(1-x^2)}{2\pi(1+x^2)^2},
\eenrr
while those corresponding to the logistic distribution will be
\benrr
v_{0}(x)&=&\frac{3e^{2x}+1}{3(e^x+1)^3},\quad v_{1}(x)=\frac{1}{3}\ln(1+e^x)-\frac{e^x\{x(3+e^{2x})+(1+e^x)\}}{3(1+e^x)^3},\\
v_{2}(x)&=&\frac{1}{(1+e^x)} + \left( -\frac{2}{(1+e^x)}-\frac{2xe^x}{(1+e^x)^2} \right) + R(x),
\eenrr
where $R(x):=\int_{x}^{\iny} y^2e^y(1-e^y)^2/(1+e^y)^4 dy$. It is trivial to see that $v_{1}$ and $v_{2}$ are continuous functions on $\mathbb{R}$ while continuity of $v_{2}$ directly follows from Lemma \ref{lem:Rx} below. Furthermore, we have the following lemma.
\begin{lem}\lel{lem:vi}
For the logistic and Cauchy distributions, $v_{i}(x), i=0,1,2$ are all strictly positive and continuous.
\end{lem}
\noi
Due to the previous argument, (\textbf{F.1})-(\textbf{F.3}) imply claims for $v_{0}$ and $v_{2}$ hold true while that for $v_{1}$ should be verified. The proof of the claim for $v_{1}$ is easy, and hence, we do not include it here. Skipping the proof of the lemma, however, doesn't diminish its importance. The validity of all subsequential findings such as nonsingularity of $\Gamma_{t}$ and the existence of solutions to (\ref{eq:teststat}) heavily depends on Lemma \ref{lem:vi}: see, e.g., Lemma \ref{lem:ci}. Figure \ref{fig:vi} serves to  graphically sustain the lemma.
\begin{figure}[h]
\centering
\includegraphics[width=0.75\textwidth]{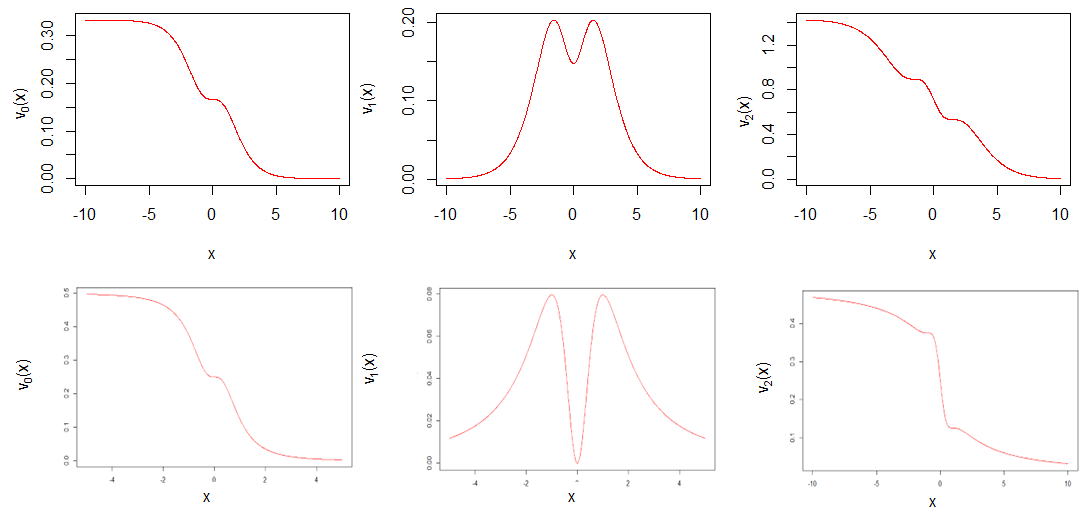}
\caption{Graphs of $v_{0}(x)$ (left), $v_{1}(x)$ (middle), and $v_{2}(x)$ (right) of the logistic (top) and Cauchy (bottom) distributions.}\lel{fig:vi}
\end{figure}

Note that the absence of a closed-form expression for $R(x)$ brings a consequential absence of a closed-form for $v_{2}(x)$, which subsequently renders the computation of $\Gamma^{-1}$ more challenging. However, $R(x)$ is bounded and decays to 0 fast as $x$ goes to $\iny$ while it converges to the finite value $(\approx 2.43)$ as $x$ goes to $-\iny$: see Figure \ref{fig:Rx}. Motivated by this fact and the following lemma, a numerical approximation to $R(x)$ will, therefore, be used to calculate the inverse of $\G$ for the logistic distribution.
\begin{lem}\lel{lem:Rx}
$R(x)$ is continuous on $\mathbb{R}$, and
\benn
\sup_{x\in\R}|R(x)|<\iny.
\eenn
Also, $R(x)$ converges to 0 as $x$ approaches to $\iny$, i.e.,
\benn
\lim_{x\rightarrow\iny}R(x)=0.
\eenn
\end{lem}
\begin{proof}
Let $r(y):=2y^2e^y(1-e^y)^2/(1+e^y)^4$ and $a(y):=y^2e^{-y}$. Note that both are continuous functions which take a positive value at all $y\in \R$; continuity of $R$, therefore, directly follows from continuity of $r$. Also, we have
\benn
\int_{x}^{\iny}a(y)dy = x^2e^{-x}+2e^{-x}(x+1)< \iny
\eenn
for all $x\in \R$. From the convergence of $e^{-x}$, $x^2e^{-x}$ and $xe^{-x}$ to 0 as $x$ goes to $\iny$, the above implicit integral also converges to 0 as $x$ goes to $\iny$. Together with the convergence of the implicit integral, the convergence of $r(y)/a(y)$ to 1 as $y$ approaches to $\iny$ implies
\benn
R(x)=\int_{x}^{\iny}r(y)dy < \iny,
\eenn
for all $x\in\R$ by the limit comparison test, thereby completing the proof of the second claim. Note that $r(y)$ is symmetric around 0, and hence,
\benrr
\lim_{x\rightarrow-\iny}R(x) &=& 2R(0)-\lim_{x\rightarrow\iny}R(x) <\iny
\eenrr
which, in conjunction with the continuity of $R$ and the second claim, completes the proof of the first claim.
\end{proof}
\begin{rem}
Figure \ref{fig:Rx} corroborates Lemma \ref{lem:Rx}.
\end{rem}
\begin{figure}[h]
\centering
\includegraphics[width=0.75\textwidth]{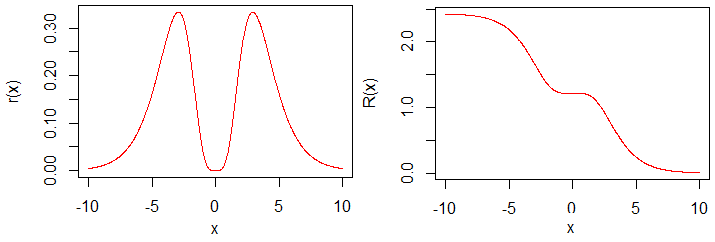}
\caption{Graphs of $r(x)$ (left) and $R(x)$ (right) over [-10, 10].}\lel{fig:Rx}
\end{figure}
Next, define
\benrr
c_{0}(x)&:=&x^{2}v_{0}(x)-2xv_{1}(x)+v_{2}(x),\\ c_{1}(x)&:=&(1-F(x))-c_{2}^{-1}(x)c_{0}(x)f^{2}(x),\\ c_{2}(x)&:=&v_{0}(x)v_{2}(x)-v_{1}^{2}(x).
\eenrr
\noi
\begin{lem}\lel{lem:c1}
For a given distribution $F$, $\Gamma_{t}$ with $t=F(x)$ is nonsingular if and only if $c_{1}(x)\neq 0$ for all $x\in \mathbb{R}$.
\end{lem}
\begin{rem}
Whether the KMT is applicable for a given distribution can be determined through checking $c_{1}\neq 0$. In other words, the distribution $F$ is Khmaladze-transformable if and only if $c_{1}\neq 0$.
\end{rem}

\begin{proof}
With $t=F(x)$, $\Gamma_{t}$ in (\ref{eq:Gamma}) can be partitioned into four $2\times2$ blocks
\benn
\Gamma_{t}= \left(
    \begin{array}{cc}
      \G_{11} & \G_{12} \\
      \G_{21} & \G_{22} \\
    \end{array}.
   \right),
\eenn
where
\benrr
\Gamma_{11} &=& 1-F(x),\quad \Gamma_{12} = (f(x), xf(x))',\\
\Gamma_{21} &=& \left(
    \begin{array}{c}
      f(x) \\
      xf(x)\\
    \end{array}
   \right),  \quad\Gamma_{22} = \left(
    \begin{array}{cc}
      v_{0}(x) & v_{1}(x) \\
      v_{1}(x) & v_{2}(x) \\
    \end{array}
   \right).
\eenrr
Then, the inverse of of $\G_{t}$ can be written as
\benn
\G_{t}^{-1} = \left(
    \begin{array}{ll}
      \Delta_{11} & \quad -\Delta_{11}\G_{12}\G_{22}^{-1} \\
      -\G_{22}^{-1}\G_{21}\Delta_{11} & \quad  \G_{22}^{-1}+\G_{22}^{-1}\G_{21}\Delta_{11}\G_{12}\G_{22}^{-1} \\
    \end{array}
   \right)
\eenn
where $\Delta_{11} = (\G_{11}-\G_{12}\G_{22}^{-1}\G_{21})^{-1}$. Clearly, the existence of the inverse of $\G_{t}$ hinges on  $\Delta_{11}\neq 0$ and non-singularity of $\G_{22}$. Note that $c_{1}=\Delta_{11}^{-1}$ and $c_{2}$ is the determinant of $\G_{t}$, and hence, $c_{i}(x)\neq 0, i=1,2$ for all $x\in \mathbb{R}$ guarantees the existence of $\G_{t}^{-1}$. Note that $c_{2}(x)>0$ for all $x\in \mathbb{R}$ follows from Cauchy-Schwartz inequality, thereby completing the proof of the lemma.
\end{proof}

After ascertaining the nonsingularity of $\Gamma_{t}$ (i.e., $c_{1}(x)\neq 0$ for all $x\in \R$), it can be shown that the fact that $c_{i}$'s are positive and continuous functions guarantees the existence of a solution to (\ref{eq:teststat}), which inevitably and intricately articulates with the Lemma \ref{lem:vi}. Kim (2020) proved the claims for $c_{i}$'s hold true for a normal distribution only, which is a definite limitation of the findings. We shall extend his findings to other distributions.
\begin{lem}\lel{lem:ci}
For the logistic and Cauchy distributions, corresponding $c_{i}(x),\,i=0,1,2$ are strictly positive and continuous functions for all $x\in\R$. Consequently, $\Gamma_{t}$'s of both distributions are nonsingular, and hence, they are Khmaladze-transformable.
\end{lem}
\begin{rem}
Unlike the logistic and Cauchy distributions, the exponential distribution has $c_{1}(x)\equiv0$ for all $x\in \mathbb{R}$, and hence, it is not Khmaladze-transformable at all.
\end{rem}
\begin{rem}\lel{strategy}
When Gordon (1941), Birnbaum (1942), Gasull and Utzet (2014), and Baricz (2008) sought the lower bound for the Mill's ratio of a normal distribution and needed to prove a function is positive, they used a basic calculus: its monotonicity and convergence to zero as $x$ goes to $-\iny$ or $\iny$. When the necessity to prove the same claim arises, we will adopt the same strategy.
\end{rem}
\begin{proof}
Observe that the claims for $c_{0}$ and $c_{2}$ directly follow from Lemma \ref{lem:vi} while one for $c_{1}$ should be carefully examined. The claim for $c_{2}$ follows from the Cauchy-Schwarz inequality and continuity of $v_{i}$. Note that $c_{0}(x)=v_{0}(x)\big\{x-v_{1}(x)/v_{0}(x)\big\}^2+c_{2}(x)/v_{0}(x)$, and hence, $c_{0}(x)>0$ for all $x\in\R$ follows from $c_{2}(x)>0$ and $v_{0}(x)>0$ for all $x\in\R$ while the continuity of $c_{0}$ is again a direct result of the continuity of $v_{i}$'s.

Now consider $c_{1}$. Here, we provide a proof of the claim for the logistic distribution only; the proof for the Cauchy case is almost the same but much simpler and easier. Continuity of $c_{1}$ is straightforward. Note that $c_{2}>0$ vouches for
\ben\lel{eq:mill}
\frac{(1-F(x))}{f(x)} >\frac{c_{0}(x)}{c_{2}(x)}f(x),
\een
where the left-hand side of the above inequality is the Mill's ratio. Therefore, proving the claim for $c_{1}$ will provide the lower bound for the Mill's ratio of the logistic distribution.

Recall $f(x)=e^{-x}/(1+e^{-x})^2$ and $F(x)=1/(1+e^{-x})$ of the logistic distribution. Using $(1-F(x))/f(x)=1/F(x)$, rewrite $c_{1}(x)$ as
\benrr
%c_{1}(x)&=& \frac{c_{2}(x)(1-F(x))-c_{0}(x)f^{2}(x)}{c_{2}(x)}\\
c_{1}(x)        &=& \frac{f(x)\{c_{2}(x)-c_{0}(x)f(x)F(x)\}}{c_{2}(x)F(x)}.
\eenrr
Together with the fact that $c_{2}(x)>0$ and $F(x)>0$ for all $x\in\R$, continuities of functions in both denominator and numerator of the above equation vouch for continuity of $c_{1}$.

Define $h(x):=c_{2}(x)-c_{0}(x)f(x)F(x)$. In conjunction with $c_{2}(x)>0$, $f(x)>0$, and $F(x)>0$ for all $x\in\R$, $c_{1}(x)>0$ is, thus, equivalent to $h(x)>0$. To show that $h(x)$ is a positive function, we shall show that $h(x)$ is a continuous, decreasing function, and $h(x)$ converges to 0 as $x$ approaches to $\iny$. To that end, it will be shown that $h'(x)<0$ for all $x\in\R$. Note that
\benrr
h'(x) &=& -(1+e^{x})^{-2}f(x)\Big\{ v_{2}(x)-2(1+x+e^{x})v_{1}(x)\\
      & &\qquad +(1+x+e^{x})^{2}v_{0}(x)-e^{x}(1+e^{x})f(x)\Big\}\\
      &=&-(1+e^{x})^{-2}f(x)j(x),
\eenrr
where $j(x):=v_{2}(x)-2(1+x+e^{x})v_{1}(x)+(1+x+e^{x})^{2}v_{0}(x)-e^{x}(1+e^{x})f(x)$. Then, the claim for $h'(x)$ can be proven by showing $j(x)>0$ for all $x\in\R$. To show $j(x)>0$, we will use the same strategy, that is, we will show that $j(x)$ is a decreasing function and converges to 0 as $x$ approaches to $\iny$. The convergence of $j(x)$ to 0 is clear. Observe that $j'(x)= -2k(x)/( 3(1+e^{x})^{3} )$ where
\benrr
k(x)&:=&(1+e^{x})^{4}\Big\{\log(1+e^{x})-(x-3)e^{4x}-(x-2)e^{3x}\\
    & &\qquad-(6x+2)e^{2x}-(3x+2)e^{x}-(x+1) \Big\}.
\eenrr
Thus, showing $k(x)>0$ for all $x\in\R$ will complete the proof of the claim. Rewrite $k(x)$ as
\benn
k(x) = \Big(\log(1+e^{x})-x\Big)(1+e^{x})^{4}+(e^{x}-1)(e^{x}+1)(3e^{2x}+2e^{x}+1)+x(e^{x}+3e^{3x}).
\eenn
Note that the second term of the righthand side of the above equation is always greater than 0; hence, we consider the first and third terms only. Expand the sum of the first and third terms and rewrite it in the descending order of $e^{x}$. Let $\kappa(x):=\log(1+e^{x})-x$; it is plain to see that $\kappa(x)>0$ for all $x\in \R$, and hence, only thing left to complete the proof is to show that $4\kappa(x)+x$ and $4\kappa(x)+3x$ -- coefficients of $e^{x}$ and $e^{3x}$, respectively -- are always greater than 0, that is,
\benn
4\kappa(x)+x>0, \textrm{ and }\,\, 4\kappa(x)+3x>0, \textrm{ for all }x<0.
\eenn
Note that $4\kappa(x)+3x$ is convex and attains its global minimum ($4\log(4/3)-1/3>0$) at $x=-\log3$. Since $4\kappa(x)+x>4\kappa(x)+3x$ for $x<0$, the above claim holds, which implies that $j(x)$ is a decreasing function, thereby completing the proof of the claim for $c_{1}$.
\end{proof}
\begin{figure}[h]
\centering
\includegraphics[width=0.75\textwidth]{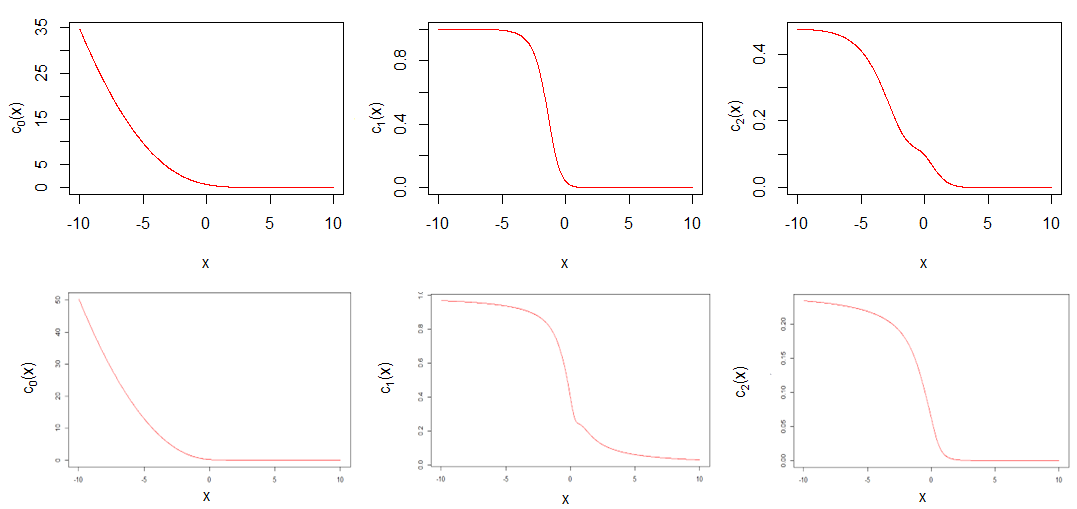}
\caption{Graphs of $c_{0}(x)$ (left), $c_{1}(x)$ (middle) and $c_{2}(x)$ (right) of the logistic (Top) and Cauchy (bottom) distributions.}\lel{fig:ci}
\end{figure}
\noi
Figure \ref{fig:ci} shows graphs of $c_{i}(x), i=0,1,2$ for the logistic and Cauchy distributions, thereby empirically proving Lemma \ref{lem:ci}. Consider the GOF test for the Gumbel distribution that also belongs to a location-scale family. Note that corresponding $v_{i}$'s and $c_{i}$'s do not have closed-form expressions, and hence, direct proof for the Khmaladze-transformability of the Gumbel distribution -- i.e., $c_{1}(x)\neq 0$ -- is not a viable option. However, as described in Figure \ref{fig:ci}, $c_{1}(x)$ of the Gumbel distribution can be empirically shown to be a strictly positive and continuous function. The same argument holds true for the Rayleigh distribution.

Note that Lemma \ref{lem:ci} provides useful lower bounds of the Mill's ratio for the logistic and Cauchy distributions.
\begin{corollary}
Consider a distribution whose density and distribution functions are $f$ and $F$, respectively. Let $\rho(x):=(1-F(x))/f(x)$ denote the Mill's ratio of the given distribution. Then, $c_{0}(x)f(x)/c_{2}(x)$ serves as a lower bound for its Mill's ratio. Especially, the lower bound of $\rho(x)$ for the Cauchy distribution is
\benn
2(1+x^2)^{-1}[\pi(0.5-\arctan(x))(1+x^2)+x]^{-1}
\eenn
while that for the logistic distribution is $\gamma_{1}(x)/\gamma_{2}(x)$ where
\benrr
\gamma_{1}(x)&=&3e^{x}(1+e^{x})(1+6e^{x}+3e^{2x}+2e^{3x})x^{2}-12e^{2x}(1+e^{x})^{2}x\\
&&\quad -6xe^{x}(1+e^{x})^{4}\ln(1+e^{x})-9e^{x}(1+e^{x})^{3}\{1+(1+e^{x})R(x)\},\\
\gamma_{2}(x)&=&3(x^{2}+3x^{2}e^{2x})(1+e^{x})^{3}R(x)-(1+e^{x})^{6}\left\{\ln(1+e^{x})\right\}^{2}\\
&&\quad + 2\ln(1+e^{x})\left\{ e^{x}(1+e^{x})(3+e^{2x})x+e^{x}(1+e^{x})^{4} \right\}\\
&&\quad -6(1+e^{x})e^{2x}(3e^{x}+1)x^{3}
 -(3+6e^{x}+21e^{2x}+18e^{3x} \\
&&\quad   +15e^{4x}+e^{6x})x^{2} -2e^{2x}(1+e^{x})(3+e^{2x})x.
\eenrr
\end{corollary}
\begin{proof}
Plugging $c_{0}(x)$, $c_{2}(x)$, and $f(x)$ into (\ref{eq:mill}) yields the desired result.
\end{proof}

Establishing the nonsingularity of $\G_{t}$, we proceed to obtain $\G_{t}^{-1}$ which is the most essential part of the computation of $\cT$. Using the inverse formula of $2\times 2$ four blocks, $\G_{t}^{-1}$ with $t=F(x)$ can be expressed as
\benn
\G_{F(x)}^{-1}\\
= \left(
     \begin{array}{ccc}
     B_{11}(x) & B_{12}(x) & B_{13}(x) \\
     B_{21}(x) & B_{22}(x) & B_{23}(x) \\
     B_{31}(x) & B_{32}(x) & B_{33}(x) \\
      \end{array}
      \right),
\eenn
where
\benrr
B_{11}(x) &=& 1/c_{1}(x),\quad B_{12}(x)=B_{21}(x)= -f(x)(v_{2}(x)-xv_{1}(x))/(c_{1}(x)c_{2}(x)),\\
B_{13}(x) &=& B_{31}(x)= -f(x)(xv_{0}(x) - v_{1}(x))/(c_{1}(x)c_{2}(x)),\\
B_{22}(x) &=& \Big\{c_{1}(x)c_{2}(x)v_{2}(x)+f^{2}(x)(xv_{1}(x)-v_{2}(x))^{2}\Big\}/(c_{1}(x)c_{2}^{2}(x)),\\
B_{23}(x) &=& B_{32}(x)= -\Big\{f^{2}(x)(xv_{1}(x)-v_{2}(x))(xv_{0}(x)-v_{1}(x))\\
&&\qquad  +c_{1}(x)c_{2}(x)v_{1}(x)\Big\}/(c_{1}(x)c_{2}^{2}(x)),\\
B_{33}(x) &=& \Big\{c_{1}(x)c_{2}(x)v_{0}(x)+f^{2}(x)(xv_{0}(x)-v_{1}(x))^{2}\Big\}/(c_{1}(x)c_{2}^{2}(x)).
\eenrr
\noi
Using these equations, $\boldsymbol{\ell}(\wh{X}_{i})'\G_{F(x)}^{-1}\boldsymbol{\ell}(x)f(x)$ can be rewritten as
\benn
\boldsymbol{\ell}(\wh{X}_{i})'\G_{F(x)}^{-1}\boldsymbol{\ell}(x)f(x)= s_{1}(x) + \phi(\wh{X}_{i})s_{2}(x) + (\wh{X}_{i}\phi(\wh{X}_{i})-1)s_{3}(x)
\eenn
where
\benr
s_{1}(x) &=& f(x)\Big\{ B_{11}(x)+B_{12}(x)\phi(x) + B_{13}(x)(x\phi(x)-1) \Big\},\lel{eq:s1}\\
s_{2}(x) &=&  f(x)\Big\{ B_{21}(x)+B_{22}(x)\phi(x) + B_{23}(x)(x\phi(x)-1) \Big\},\nonumber\\
s_{3}(x) &=&  f(x)\Big\{ B_{31}(x)+B_{32}(x)\phi(x) + B_{33}(x)(x\phi(x)-1) \Big\}.\nonumber
\eenr
\noi
Recall that $c_{2}(x)$ is strictly positive for all $x\in \R$. Thus, $c_{1}(x)\neq 0$ for all $x$ implies that $B_{ij}$'s for $1\leq i, j\leq 3$ are all continuous functions, thereby demonstrating that $s_{i}$'s are also continuous functions. Let $G_{i}(y):= \int_{-\iny}^{y}\boldsymbol{\ell}(\wh{X}_{i})'\G_{F(x)}^{-1}\boldsymbol{\ell}(x)f(x)\,dx$ and observe that
\benrr
G_{i}(y) &=& \int_{-\iny}^{y}s_{1}(x)\,dx + \phi(\wh{X}_{i})\int_{-\iny}^{y}s_{2}(x)\,dx + (\wh{X}_{i}\phi(\wh{X}_{i})-1)\int_{-\iny}^{y}s_{3}(x)\,dx,\\
&=& S_{1}(y) + \phi(\wh{X}_{i})S_{2}(y)+(\wh{X}_{i}\phi(\wh{X}_{i})-1)S_{3}(y), \quad say.
\eenrr

\begin{lem}\lel{lemma:Gi}
For the logistic and Cauchy distributions, $S_{j}(y),\,\, j=1,2,3$ converge to 0 as $y$ approaches to $-\iny$. Consequently, the same result will hold for $G_{i}(y),\,i=1,2,...,n$.
\end{lem}
\begin{proof}
Here we prove the claim for $S_{1}$ only since the exactly same argument can be applied for $S_{2}$ and $S_{3}$.
To begin with, rewrite
\benn
\lim_{x\rightarrow-\iny} S_{1}(x)=\lim_{x\rightarrow\iny}\int_{x}^{\iny} s_{1}(-y)dy,
\eenn
which implies integrability of $s_{1}(-y)$ will complete the proof. As shown in the previous paragraph, $s_{1}(-y)$ is the sum of three terms: $f(y)B_{11}(-y)$, $f(y)B_{12}^{11}(-y)\phi(-y)$, and $f(y)B_{12}^{12}(-y)(-y\phi(-y)-1)$. Consider the Logistic distribution first. Note that for large enough $x\in\R$, $v_{0}(y)\approx e^{-y}$, $v_{1}(y)\approx y$, $v_{2}(y)\approx e^{-y}$. Using this fact, it can be easily shown that $B_{11}(-y)f(y)\approx y^{-2}$. Then, the integrability of $y^{-2}$ and the limit comparison test imply the improper integral of $f(y)B_{11}(-y)$ is finite, i.e,
\benn
\lim_{x\rightarrow\iny} \int_{x}^{\iny} f(y)B_{11}(-y) dy<\iny.
\eenn
The finite improper integrals of other terms of $s_{1}(-y)$ can be shown in the similar way, which implies integrability of $s_{1}(-y)$, thereby proving the claim for $S_{1}$ of the Logistic distribution.

Next, consider the Cauchy distribution. Note that $c_{1}(x)$ can be rewritten as
\benn
c_{1}(x)=(1-F(x))-\frac{2}{\pi^2(1-F(x))(1+x^2)+\pi x}.
\eenn
Using these simplified $c_{1}$, we can derive the integrability of $s_{1}$ by demonstrating the integrability of each term in the right-hand side of the equation in (\ref{eq:s1}). Consider, for example, $f(x)B_{11}(x)$. Since $f(x)B_{11}(x)\approx 1/(1+x^2)$ for $x$ large enough, its integrability is obvious. Similarly, we have
$f(x)B_{12}(x)\phi(x)\approx x^{2}/(1+x^2)^{3}$ and $f(x)B_{13}(x)(x\phi(x)-1)\approx x^{6}/(1+x^2)^{4}$, which in turn implies the integrability of $s_{1}$, thereby demonstrating the convergence of $S_{1}(x)$ to 0 as $x\rightarrow -\iny$.

\end{proof}
Now we are ready to state the main result of this article.
\begin{theorem}\lel{thm:1}
Assume we observe a finite sample of i.i.d.\,\, $X_{1},...,X_{n}$. Consider a GOF test in (\ref{eq:1}) and a optimization problem in (\ref{eq:teststat}). Assume (\textbf{F.1})-(\textbf{F.3}), (\textbf{E}), $c_{1}(x)\neq 0$ for all $x\in \R$, and $S_{j}(y),\,j=1,2,3$ converge to zero as $y$ goes to $-\iny$. Then, a finite solution to (\ref{eq:teststat}) almost surely exists.
\end{theorem}
\begin{rem}
Theorem \ref{thm:1} implies the GOF tests for both logistic and Cauchy distributions have the finite solution: see also Figure \ref{fig:Unz}. When testing for other distributions -- e.g., a Gumbel distribution -- which do not have closed-from expressions for $v_{i}$'s and $c_{i}$'s, the practitioner can empirically check whether the assumptions of the theorem are met.
\end{rem}
\begin{proof}
As shown in Kim (2020), $\mathcal{U}_{n}$ can be simplified and rewritten as
\benrr
\mathcal{U}_{n}(z)&=& \left\{
           \begin{array}{ll}
             -n^{-1/2}\sum_{k=1}^{n}G_{k}(z) & \hbox{if $z< \wh{X}_1$;} \\
             n^{1/2} - n^{-1/2}\sum_{k=1}^{n}G_{k}(\wh{X}_k), & \hbox{if $z\ge \wh{X}_n$,}\\
           \end{array}
         \right.
\eenrr
while
\benn
\mathcal{U}_{n}(z)= -n^{-1/2}\sum_{k=i+1}^{n}G_{k}(z) +n^{-1/2}\sum_{l=1}^{i}(1 - G_{l}(\wh{X}_l)),
\eenn
when $z\in [\wh{X}_{i}, \wh{X}_{i+1})$ for $i=1,2,...,n-1$. Consider $(n+1)$ intervals: $(-\iny, \wh{X}_{1})$, $[\wh{X}_{1}, \wh{X}_{2})$,..., $[\wh{X}_{n-1}, \wh{X}_{n})$, $[\wh{X}_{n}, \iny)$. Over these intervals, $\mathcal{U}_{n}(z)$ will be shown to; (1) converge to 0 over the first interval; (2) take a constant over the last interval; and (3) be piecewise continuous over other intervals.

The last feature is clear. Continuity of $G_{i}$ directly follows from the continuity of $s_{j}$'s consequential upon the assumption $c_{1}\neq 0$, thereby implying the second feature. The assumption of convergence of $S_{j}$'s to 0 implies the convergence of $G_{k}(z),\,k=1,..,n$ to 0 as $z\rightarrow-\iny$, and hence the convergence of $\mathcal{U}_{n}(z)$ to 0 follows. Thus, the piecewise continuity of $\mathcal{U}_{n}(z)$ over the bounded intervals together with either converging to zero or
being constant over unbounded intervals proves the claim, thereby completing the proof of the theorem.
\end{proof}

\begin{figure}[h]
\centering
\includegraphics[width=0.75\textwidth]{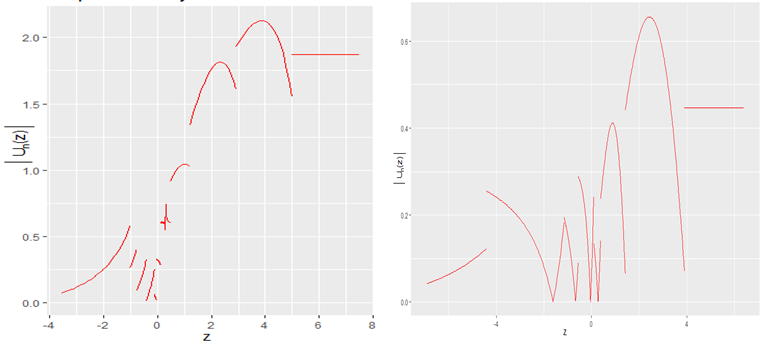}
\caption{Graphs of $|\mathcal{U}_{n}(z)|$ of the logistic (left) and Cauchy (right) distributions.}\lel{fig:Unz}
\end{figure}
Now, we conclude this section with discussion of the modified KMT test. Let $\mathcal{H}:=\{\textrm{function }\,h:\R\rightarrow \R\}$, that is, a set of functions on $\R$. Consider $\varphi:\mathcal{H}\rightarrow \R$ where
\benn
\varphi(h) = \sup_{x\in \R} h(x) - \inf_{x\in \R}h(x).
\eenn
A uniform metric for any $h_{1}, h_{2}\in \mathcal{H}$ is defined as
\benn
\rho(h_{1}, h_{2}) = \sup_{x\in \R}|h_{1}(x)- h_{2}(x)|.
\eenn
Observe that $\varphi$ is continuous in a uniform metric. Recall the modified KMT test statistic $\cT^{M}$ in (\ref{eq:TM}). Thus, the ADF convergence of $\cT^{M}$ to
\benn
\sup_{x\in\R}\mathcal{B}(F(x))-\inf_{x\in\R}\mathcal{B}(F(x))
\eenn
in distribution immediately follows from the continuity of $\varphi$ where $\mathcal{B}$ is a standard Brownian motion. We conjecture that that $\cT^{M}$ will detect any deviations from the null distribution better than $\cT$. To put it another way, $\cT$ detects only the biggest deviation regardless of whether it is positive or negative while $\cT^{M}$ cares the biggest positive and negative deviations together. Then, common sense suggests that this slight modification is expected to enhance the performance of the KMT test in that its power will be increased. The following lemma indeed verifies that our conjecture is true.
\begin{lem}\lel{lem:power}
Recall the GOF test in (\ref{eq:1}). Let the KMT and modified KMT test statistics are as in (\ref{eq:teststat}) and (\ref{eq:TM}), respectively. Then, the modified KMT test always has a better power than the original KMT test.
\end{lem}
\begin{proof}
Consider $\{\cT>\kappa\}$ and $\{\cT^{M}>\kappa\}$ for any $\kappa\geq 0$. Let $\phi(h):=\sup_{x\in \R}|h(x)|$. Then, $\cT$ and $\cT^{M}$ can be rewritten as $\phi(\mathcal{U}_{n})$ and $\varphi(\mathcal{U}_{n})$, respectively. Let $a\vee b:=\max(a,b)$ for $a, b\in R$ and $\|h\|_{\iny}=\sup_{x\in \R}|h(x)|$ for $h\in \mathcal{H}$. Using $ \|h\|_{\iny}=\sup_{x\in \R}h(x) \vee \sup_{x\in \R}-h(x)$, we obtain $\varphi(h)\geq \phi(h)$ for all $h\in \mathcal{H}$.

Let $\beta_{1}(\kappa)$ and $\beta_{2}(\kappa)$ denote the statistical powers of the original and modified KMT tests, respectively. Then,
\benn
\beta_{1}(\kappa)=\mathbb{P}(\cT>\kappa\,|\,\textrm{$H_{0}$ is false})\leq \mathbb{P}(\cT^{M}>\kappa\,|\,\textrm{$H_{0}$ is false})=\beta_{2}(\kappa),
\eenn
where the inequality follows from $\varphi\geq \phi$, thereby competing the proof of the lemma.
\end{proof}
\begin{rem}
As mentioned earlier, the result in Theorem \ref{thm:1} still holds for the modified KMT test, i.e., there almost surely exists $z_{*}\iny$ which is a solution to (\ref{eq:TM}) since solving optimization still exploits the features of $\mathcal{U}_{n}$: convergence to 0, piecewise continuity, and constant over intervals determined by sample observations.
\end{rem}

\section{Simulation studies}\lel{sec:simulation}
In this section, we will compare the original and modified KMT tests for the logistic and Cauchy distributions. For the comparison purpose, empirical levels and statistical powers of two tests will be examined. Finally, we conclude this section by presenting an example of the application of the KMT test to real data; this example will attest to its efficiency proven in the simulation studies. In the sequel, $\alpha$ denotes the significance level.
\subsection{Asymptotic distributions}
Recall $\phi$ and $\varphi$ from the previous section. From the convergence of $\cT$ and $\cT^{M}$ to $\phi(\mathcal{B}\circ F)$ and $\varphi(\mathcal{B}\circ F)$, respectively, asymptotic critical values for both statistics can be obtained by generating random samples of corresponding asymptotic distributions.
To this end, we generate a Brownian motion over $[0,1]$ and find $\phi(\cdot)$ and $\varphi(\cdot)$. Then, we iterate this procedure 10,000 times, thereby having random samples of 10,000 observations of $\phi(\mathcal{B}\circ F)$ and $\varphi(\mathcal{B}\circ F)$. Finally, we compute the $(1-\alpha)\times 100$ percentiles from the samples, which play roles as asymptotic critical values for the original and modified KMT tests. From the simulations, we have 2.231 and 2.478 for the critical values of $\alpha=0.05$ for both the original and modified tests, respectively: note that 2.231 of the original test corresponding to $\alpha=0.05$ closely accords with findings in other works: e.g., see Khmaladze and Koul (2004).
\begin{figure}[h]
\centering
\includegraphics[width=0.75\textwidth]{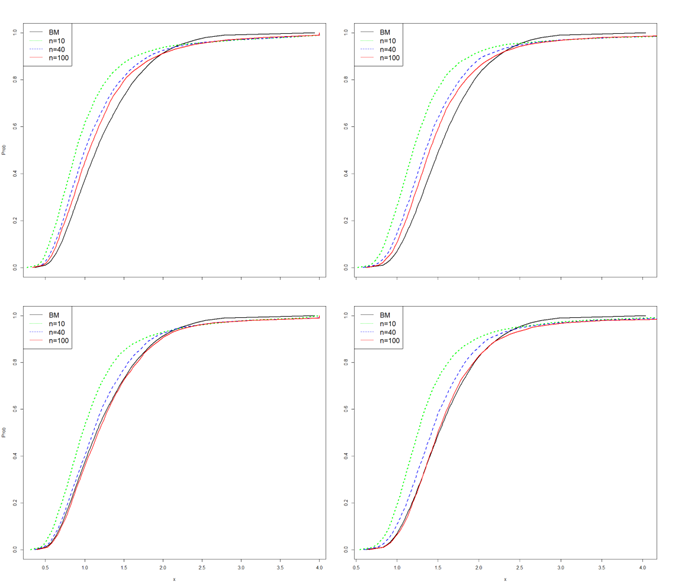}
\caption{Convergence of $\cT$ (left) and $\cT^{M}$ (right) of the logistic (top) and Cauchy (bottom) distributions.}\lel{fig:conv}
\end{figure}
Figure \ref{fig:conv} displays the convergence of $\cT$ and $\cT^{M}$ of the logistic and Cauchy distributions when the sample size $n$'s are 10 (green), 40 (blue), and 100 (red). As described in the figure, both quickly converge to their asymptotic distributions (black) as $n$ increases.
\subsection{Empirical level and statistical power}
In this section, we will compute empirical levels and statistical powers of two KMT tests for the logistic and Cauchy distributions with various sample sizes (50,70,90,100). For each distribution, three alternative distributions from a location-scale family (normal, logistic, Cauchy, and Laplace except itself) will be tried when the statistical power being computed while the distribution itself will be tried for the empirical level. When we generate a random sample from hypothesized distributions of a location-scale family, we will use 1 and 2 for the location (=$\mu$) and scale (=$\sigma$) parameters, respectively. When the necessity to estimate $\mu$ and $\sigma$ arises for the two KMT tests (e.g., $\wh{X}_{i}$ in computing $\cU_{n}$), we will use the ML estimators.

For testing for a given distribution, test statistics ($\cT$ and $\cT^{M}$) are computed based on the sample of random variables generated from the hypothesized distributions and compared with 2.231 (the critical values for $\alpha=0.05$) to reject the null hypothesis or not. Then, we again repeat this procedure 10,000 times and compute empirical levels and statistical powers of the original and modified KMT tests through dividing the total number of the rejections by 10,000.
\begin{table}[h]
\centering
\begin{tabular}{c c cc c c}
  \hline
  &      &  Logistic   &           & Cauchy       &         \\
  \cline{3-6}
  & $n$  & $\cT$ & $\cT^{M}$ & $\cT$ & $\cT^{M}$ \\
  \hline
\multirow{5}{*}{Normal}
  &50&0.007 &0.017 &0.063 &0.482 \\
 &70&0.011 &0.025  &0.345 &0.903 \\
 &90&0.021 &0.050  &0.740 &0.992 \\
 &100&0.027 &0.054 &0.865 &0.999 \\
\hline
\multirow{5}{*}{Logistic}
 &50&0.033   &0.030 &0.022 &0.241 \\
  &70&0.032  &0.030 &0.201 &0.665 \\
  &90&0.034  &0.036 &0.508 &0.923 \\
  &100&0.043 &0.043 &0.632  &0.966 \\
 \hline
\multirow{5}{*}{Cauchy}
  &50&0.887  &0.935 &0.055 &0.056\\
  &70&0.961  &0.987 &0.063 &0.066\\
  &90&0.988  &0.998 &0.063 &0.064\\
  &100&0.992 &1.000 &0.051  &0.047\\
 \hline
\multirow{5}{*}{Laplace}
  &50&0.127  &0.127 &0.011 &0.051\\
  &70&0.147  &0.152 &0.042 &0.185\\
  &90&0.202  &0.207 &0.183 &0.437\\
  &100&0.219  &0.229&0.275 &0.554\\
 \hline
\end{tabular}
\caption{Empirical levels and statistical powers of tests for the logistic and Cauchy distributions}\lel{tbl:simulation}
\end{table}
Tables \ref{tbl:simulation} reports empirical levels and statistical powers of two KMT tests when the sample size $n$ varies (50, 70, 90, and 100): the last two columns report results of the Cauchy distribution while the others report those of the logistic distribution. Consider the logistic (Cauchy) distribution. The row corresponding Logistic (Cauchy) reports empirical levels of two KMT tests while other rows report their statistical powers. The most prominent feature underlying the result is that the modified KMT test outperforms the original test in terms of the statistical power regardless of alternative distributions. For example, in case of testing for the Cauchy distribution, the modified test displays two ($n=100$) to five ($n=50$) times greater statistical power than original test when the alternative distribution is the Laplace: this result coincides exactly with Lemma \ref{lem:power}. On the other hand, the original test shows similar (or slightly better) empirical level to (than) the modified test.

\subsection{Real data}

In probability theory, testing for non-normal distributions came to the fore of attention. Consider, e.g., the logistic and Cauchy distribution. The literature abounds with articles which illustrate that the logistic distribution is extensively used in various disciplines. The most oft-quoted example is a logistic regression model for categorical data analysis which is very popular in social science: see, e.g., Balakrishinan (1992). The Cauchy distribution has also been popular since it can successfully explain some extreme behaviors of economic data such as a sudden collapse of stock price and a radical increase or decrease in the market sale of corporate companies: see Granger and Orr (1972) for more details. Besides these two distributions, applications of other distributions -- e.g., extreme value distributions -- can be easily found in many other fields. Crucially, however, research on the GOF test for non-normal distributions has not been active as much as research on its real application, to say the least. It is against this backdrop of a rarity of a GOF test for other non-normal distributions that we started this study.

As a sequel to the simulation studies, we present an example of application of the GOF tests for the Cauchy distribution to real data which is mentioned in Introduction. Now the null hypothesis is that the underlying distribution of observations from the real data is the Cauchy distribution. By implementing the KMT test for the Cauchy distribution, we make a decision: either reject or do not reject the null hypothesis.

Figure \ref{fig:Index1} shows time series of the stock market indices in Korea and the US from December 19, 2019 to May 20, 2020: Korea Composite Stock Price Index 200 (KOSPI200)\footnote{It is available at \url{http://global.krx.co.kr}.} and Dow Jones Industrial Average Index (DOWJONES)\footnote{It is available at \url{https://www.wsj.com}}.
\begin{figure}[h]
\centering
\includegraphics[width=0.5\textwidth]{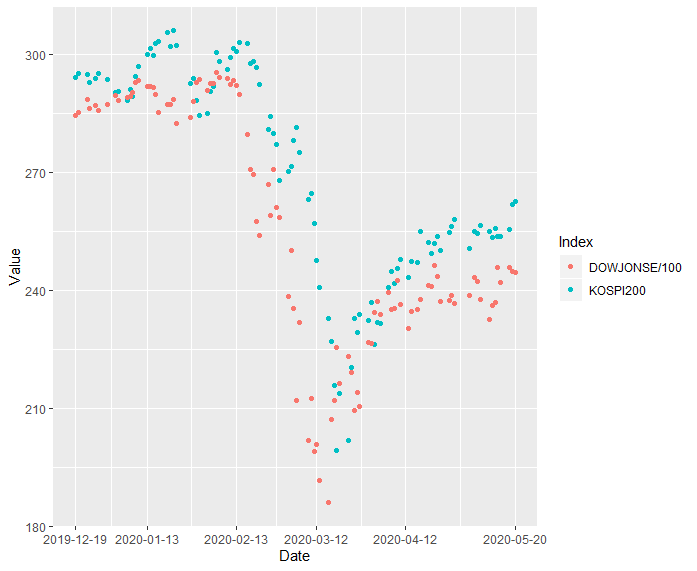}
\caption{Chart of KOSPI200 and DOWJONES}\lel{fig:Index1}
\end{figure}
For the convenience of the analysis, we scale down nominal values of the DOWJONES at a 1:100 scale while those of the KOSPI200 remain the same: for example, 284.6214 of the DOWJONES at the starting date represents 28462.14 while its KOSPI200 counterpart (294.31) represents the same value.

These two indices display the same pattern, that is, a sudden collapse -- starting from February 13, 2020 and continuing to March 12, 2020 -- followed by a rebound and rally thereafter. The sudden collapse is imputed to the shock caused by Coronavirus Disease-19. As mentioned in the introduction, the Cauchy distribution is known to provide a better fit to the time series -- especially the financial market data which entails anomalies including the sudden collapse of the stock price -- than any other distributions.
\begin{figure}[h]
\centering
\includegraphics[width=0.8\textwidth]{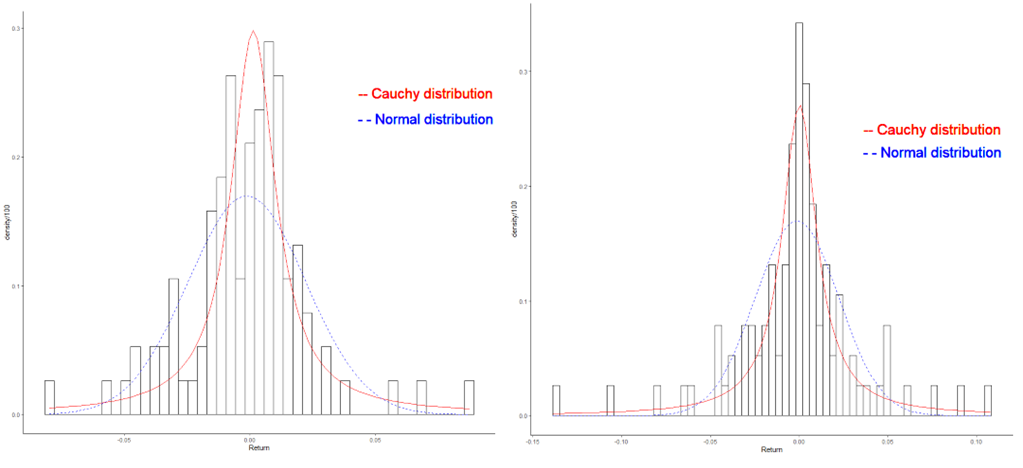}
\caption{Empirical distributions of the 100 RoR's of KOSPI200 (left) and DOWJONES (right).}\lel{fig:Index2}
\end{figure}
Figure \ref{fig:Index2} shows histograms of 100 observations of the rate of return (RoR) of two indices during the same period. The Cauchy density functions with the estimated location and scale parameters from the observation -- 0.0013 and 0.0106 for the KOSPI200 and 0.0002 and 0.0117 for the DOWJONES -- are then overlayed on the histograms. First, a quick glance reveals that a normal distribution fails to yield a decent fit to empirical distributions of both indices. Second, the KOSPI200 displays two peaks and seems to follow a bimodal distribution, which suggests the Cauchy distribution might not be the underlying distribution of the KOSPI200. On the other hand, the Cauchy distribution provides much better fit to the DOWJONES than the KOSPI200 while a normal distribution still displays a poor fit.
\begin{table}[h]
\centering
\begin{tabular}{cccc}
  \hline
        &Critical value& KOSPI200 & DOWJONES \\
  \hline
  $\cT$    &2.231 & 2.296 & 1.698 \\
  $\cT^{M}$&2.478 & 2.605 & 1.812 \\
  \hline
\end{tabular}
\caption{Test statistics and critical values of GOF tests}\lel{tbl:Index}
\end{table}
Table \ref{tbl:Index} reports the result of the original $(\cT)$ and modified $(\cT^{M})$ KMT tests based on the real data. The second column of the table reports critical values of the tests for $\alpha=0.05$ while the third and last columns report the test statistics obtained as a result of administrating those tests to the KOSPI200 and DOWJONES, respectively. When the KOSPI200 is considered, both tests reject the null hypothesis that the underlying distribution is the Cauchy distribution. The case of the DOWJONES shows a contrary result to the previous case: both tests also yield the same decision not to reject the null hypothesis. Consequently, the result shown in the table closely accords with the previous findings from Figure \ref{fig:Index2}. By yielding the same decision as the already accredited the KMT test, it is not rash for us to conclude that the modified KMT test is also qualified for the GOF test.

\section{Conclusion}
This study demonstrates that the main idea of Kim (2020) can be safely applied to the logistic and Cauchy distributions while leaving a heuristic sketch for the applicability to other non-Gaussian distributions belonging to the location-scale family such as the Weibull, Gumbel, and Frechet distributions.
The application to those distributions will, thus, form future research. Drawing all strands of our findings in the simulation studies  together, we can conclude that the modified KMT test is better than the original one since it shows the better powers for most of alternative distributions. The example of real data demonstrates the KMT test can be considered as an alternative to existing GOF tests.

\edt

outlandish, outrageous, paltry